\pdfoutput=1
\documentclass[12pt]{article}

\usepackage[margin=.99in]{geometry}

\usepackage[font=footnotesize,labelfont=bf]{caption}

\usepackage{amsmath,dsfont}
\usepackage{amssymb}
\usepackage{graphicx,color}
\usepackage{slashed,cite,xcolor}
\usepackage{hyperref}

\newcommand{\be}{\begin{equation}}
\newcommand{\ee}{\end{equation}}
\newcommand{\ba}{\begin{eqnarray}}
\newcommand{\ea}{\end{eqnarray}}

\title{{\bf \bigskip \Large{Inflation, Higher Spins and the Swampland}}\bigskip}

\author{Marco Scalisi\footnote{e-mail: {\tt marco.scalisi@kuleuven.be}} 
\bigskip\\
\bigskip\\
\small Institute for Theoretical Physics, KU Leuven, \\ \small Celestijnenlaan 200D, 3001 Leuven, Belgium \bigskip\\
\small 
}

\date{}

\begin{document}

\maketitle

\vspace{2cm}
\abstract{
\noindent We study the implications on inflation of an infinite tower of higher-spin states with masses falling exponentially at large field distances, as dictated by the Swampland Distance Conjecture. We show that the Higuchi lower bound on the mass of the tower automatically translates into an upper bound on the inflaton excursion.  Strikingly, the mere existence of {\it all} spins in the tower forbids any scalar displacement whatsoever, at arbitrarily small Hubble scales,  and it turns out therefore incompatible with inflation. A certain field excursion is allowed only if the tower has a cut-off in spin. Finally, we show that this issue is circumvented in the case of a tower of string excitations precisely because of the existence of such a cut-off, which decreases fast enough in field space.
}

\thispagestyle{empty}

\newpage

\section{Introduction}

Not all effective field theories (EFTs) admit ultraviolet completion into quantum gravity. This is probably the most famous slogan of the {\it swampland program} \cite{Vafa:2005ui,Ooguri:2006in} (see \cite{Brennan:2017rbf,Palti:2019pca} for some reviews), which aims at identifying universal constraints that quantum gravity would impose at lower, naively decoupled, energies. If correct, this approach would allow us to distinguish between healthy and pathological EFTs, where just the former would satisfy such constraints. It reserves therefore tremendous implications primarily for the understanding of the structure of the rich {\it landscape} of effective theories, which plausibly arise as low-energy limit of string theory. Certainly good news for string phenomenology, which can hope to make predictions at accessible energies, below the Planck scale, and even go beyond the standard top-down model-building approach.
 
A number of swampland criteria have been so far proposed, each with different level of evidence and predictive power \cite{Reece19}.  Among the most rigorous statements, we find the Swampland Distance Conjecture (SDC) \cite{Ooguri:2006in}. It claims that  infinite field distances are always associated with the appearance of an infinite tower of exponentially light states (see also \cite{Lust:2019zwm}). This  is intimately linked to a drop-off of the quantum gravity cut-off, which sets the scale at which the EFT breaks down. Concretely, an effective theory can only have a finite diameter of validity. This fact has collected very compelling evidence in string theory, at infinite distance regions of moduli space \cite{Grimm:2018ohb,Blumenhagen:2018nts,Lee:2018urn,Lee:2018spm,Grimm:2018cpv,Buratti:2018xjt,Gonzalo:2018guu,Corvilain:2018lgw,Font:2019cxq,Lee:2019wij} (see also \cite{Palti:2015xra,Baume:2016psm,Valenzuela:2016yny,Bielleman:2016olv,Blumenhagen:2017cxt,Palti:2017elp,Hebecker:2017lxm,Cicoli:2018tcq}). 

The phenomenological implications of the distance conjecture may therefore be strong, especially if our effective theory deals with large field distances and high energies. A typical example is inflation, where, in its simplest implementation, a scalar field traverses a certain range in order to deliver a quasi-de Sitter (dS) phase. Validity of an inflationary EFT imposes the Hubble energy scale to be always below the quantum gravity cut-off. Therefore, an exponential fall-off in field space of the latter  necessarily implies a maximum distance the inflaton may travel before the theory breaks down.  In \cite{Scalisi:2018eaz}, it has been shown that this conclusion can be made precise in terms of the tensor-to-scalar ratio measured at typical  Cosmic Microwave Background (CMB) scales (see also  \cite{Dias:2018ngv,Bravo:2019xdo} for other related works).

The features of the  tower of states usually depend on the details of the UV embedding scenario. In this letter, we would like to contemplate the possibility that this infinite tower contains {\it all} spin values.  Particles with higher-spin (HS) can in fact naturally arise in backgrounds with curvature. In this case, one can circumvent a number of severe restrictions, which applies to the flat case \cite{Weinberg:1964ew,Coleman:1967ad} (see \cite{Bekaert:2010hw} for a review), and construct consistent massless HS theories \cite{Vasiliev:1990en} (see e.g. \cite{Aros:2017ror} for some cosmological applications).  Mathematical consistency would imply that an infinite tower of fields of all spins should exist. The massive case has been studied in \cite{Singh:1974qz,Singh:1974rc} in flat space and generalized to (A)dS in \cite{Zinoviev:2001dt}. In the context of inflation, we have seen a renewed interest in this topic, given the potential phenomenological implications. HS massive particles in fact give rise to distinct signatures in cosmological correlators of the comoving curvature perturbation \cite{Arkani-Hamed:2015bza,Lee:2016vti,Kehagias:2017cym,Alexander:2019vtb}.

In the following, we discuss the non-trivial implications of having a HS tower with masses which decay  exponentially in field space, in a (quasi-)de Sitter background. We show that there is a one-to-one correspondence between a \textit{single} HS state and a specific maximum value for the inflaton range. Exceeding this value implies a violation of the Higuchi bound \cite{Higuchi:1986py} and unitarity of the theory is not preserved.  The higher the spin of the state, the smaller the allowed field distance so that the existence of {\it all} spins in the tower becomes simply incompatible with inflation.  Further, we show that the only way-out is having a cut-off in spin. This allows for some finite field excursion, in a specific range of super-Hubble masses of the spin tower.  Finally, we apply this whole argument to the tower of string states and notice that their natural cut-off (in spin and length) in de Sitter space \cite{Noumi:2019ohm,Lust:2019lmq} depends on the mass and thus allows to overcome any dire consequence for inflation. While the bound could be very dramatic for the inflationary paradigm, we are at present not aware of any specific example in perturbative string theory with a mass-independent cut-off in spin.

The outline is as follows. We start with a resume of the main properties of the SDC in Sec.~\ref{Sec:SDC}. We continue, in Sec.~\ref{Sec:spin}, by adding spin to the infinite tower and discussing the drastic implications for inflation. In Sec.~\ref{Sec:string}, we discuss the loophole in the case of the string tower. In Sec.~\ref{Sec:disc} we present our conclusions. 

\section{Swampland distance conjecture}\label{Sec:SDC}

The Swampland Distance Conjecture (SDC) \cite{Ooguri:2006in} predicts the breakdown of an effective field theory due to a tower of infinite states becoming exponentially light at large field distances with masses
\be\label{SDC}
\begin{aligned}
m=m_0\ e^{-\lambda \Delta\varphi} \qquad \text{as} \qquad\Delta\varphi\rightarrow\infty\,,
\end{aligned}
\ee
where $\lambda$ is a numerical parameter (it has been conjectured to be always of order one \cite{Klaewer:2016kiy}) and $\varphi$ parametrizes the geodesic proper distance in field space.

This breakdown happens because a description with an infinite number of light fields weakly coupled to Einstein gravity is not possible. The quantum gravity cut-off $\Lambda_{\text{QG}}$ will then experience a similar exponential drop-off
\be\label{QGcutoff}
\begin{aligned}
\Lambda_{\text{QG}}=\Lambda_0\ e^{-\gamma \Delta\varphi} \qquad \text{as} \qquad\Delta\varphi\rightarrow\infty\,,
\end{aligned}
\ee
where $\Lambda_0\leq M_P$ is the original naive cut-off of the effective theory ($M_P$ being the reduced Planck mass). Moreover, it can be shown that $\gamma = \lambda/3$  if the quantum gravity cut-off $\Lambda_{\text{QG}}$ is identified with the species scale $\Lambda_{\text{S}}=M_{P} /\sqrt{\mathcal{N}}$ \cite{Grimm:2018ohb,Heidenreich:2018kpg,Hebecker:2018vxz}, which is the cut-off of an effective gravitational theory reduced in the presence of a large number $\mathcal{N}$ of species \cite{Dvali:2007wp,Dvali:2007hz}. Above this scale, gravity becomes strongly coupled and quantum effects cannot be ignored due to the increasing number of light fields.

Evidences in string theory of the SDC can be found around infinite distance singularities of the moduli space \cite{Palti:2015xra,Baume:2016psm,Valenzuela:2016yny,Bielleman:2016olv,Blumenhagen:2017cxt,Palti:2017elp,Hebecker:2017lxm,Cicoli:2018tcq,Grimm:2018ohb,Blumenhagen:2018nts,Lee:2018urn,Lee:2018spm,Grimm:2018cpv,Buratti:2018xjt,Gonzalo:2018guu,Corvilain:2018lgw,Font:2019cxq,Lee:2019wij} (e.g. large volume, large complex structure or weak coupling points). The proper field distance to reach one of such a singular points results in fact always divergent. In concrete examples, one may check that, an infinite tower of states (typical examples are KK or winding modes) becomes exponentially light while approaching this singularity. A recent investigation \cite{Lee:2019wij} has argued that, in the infinite distance limit, either a theory decompactifies (an infinite tower of light KK modes appears) or, if the dimension does not change, the theory reduces to a weakly coupled string theory.

An immediate consequence of the drop-off of $\Lambda_{\text{QG}}$, as given by eq.~\eqref{QGcutoff}, is an upper bound in field space  which sets the range of validity of the  EFT. In the context of inflation, it has been shown \cite{Scalisi:2018eaz} that one can derive a universal upper bound on the inflaton range, which depends logarithmically on the tensor-to-scalar ratio measured at CMB scales. Moderately super-Planckian distances ($\Delta\varphi\lesssim 10$ for $\gamma=1$) are allowed, given the current experimental bound  \mbox{$H<2.5\cdot10^{-5}\ M_P$}  on the separation between the Hubble and the Planck scale, provided by the latest Planck measurements \cite{Akrami:2018odb}. Multi-field scenarios may relax the bound thanks to the possibility of following non-geodesic motion \cite{Scalisi:2018eaz,Aragam:2019khr,Bravo:2019xdo}. In one of the simplest cases, one may allow for the simultaneous variation of a saxion (the radial coordinate) and an axion (the angular coordinate) \cite{Scalisi:2018eaz}. In this case, one may check that the travelled distance, before reaching the region where the EFT breaks down, is always larger compared to the radial length.

\section{SDC and higher-spin  tower}\label{Sec:spin}

In this section, we would like to contemplate the possibility that the masses of an infinite tower of HS states follow the behaviour dictated by the SDC via eq.~\eqref{SDC}. In the case of a tower containing states with spin $s>1$, in (quasi-)de Sitter space, the situation becomes more delicate than what described in the previous section. In fact, unitarity demands that mass and spin should satisfy the {\it Higuchi bound} \cite{Higuchi:1986py}
\be\label{higuchi}
\begin{aligned}
m^2>s(s-1) H^2\,,
\end{aligned}
\ee
where $H=1/R$ is the Hubble parameter and $R$ is the de Sitter radius. 

If the HS tower has masses which follow eq.~\eqref{SDC}, then there will be a point in field space beyond which the Higuchi bound eq.~\eqref{higuchi} will be violated. This implies an upper bound on the field range such as
\be
\begin{aligned}\label{bound0}
\Delta\varphi< \frac{1}{\lambda}\log\left[ \frac{m_0}{H}\frac{1}{\sqrt{s(s-1)}}\right]\,,
\end{aligned}
\ee
obtained by combining eq.~\eqref{SDC} and eq.~\eqref{higuchi}. In any realistic model of inflation, the Hubble parameter $H$ will have an explicit field dependence. However, in this specific case, we can assume it to be constant if this dependence is milder than the exponential drop-off in field space of the mass of the infinite tower of spin modes. In fact, please note that while slow-roll inflation imposes the slope of the potential to be very small (i.e. it is a quasi-de Sitter phase), the rate at which the mass of the tower decreases is lower-bounded \cite{Grimm:2018ohb}. The mass $m_0$ is assumed to be constant in this section. Depending on the specific tower, it can however show some spin-dependence but the main result will still hold, if the mass does not grow faster than linear. A specific example, where $m_0$ depends on $\sqrt{s}$, is given in the following section.

\begin{figure}[t]
\hspace{-3mm}
\begin{center}
\includegraphics[width=7.2cm]{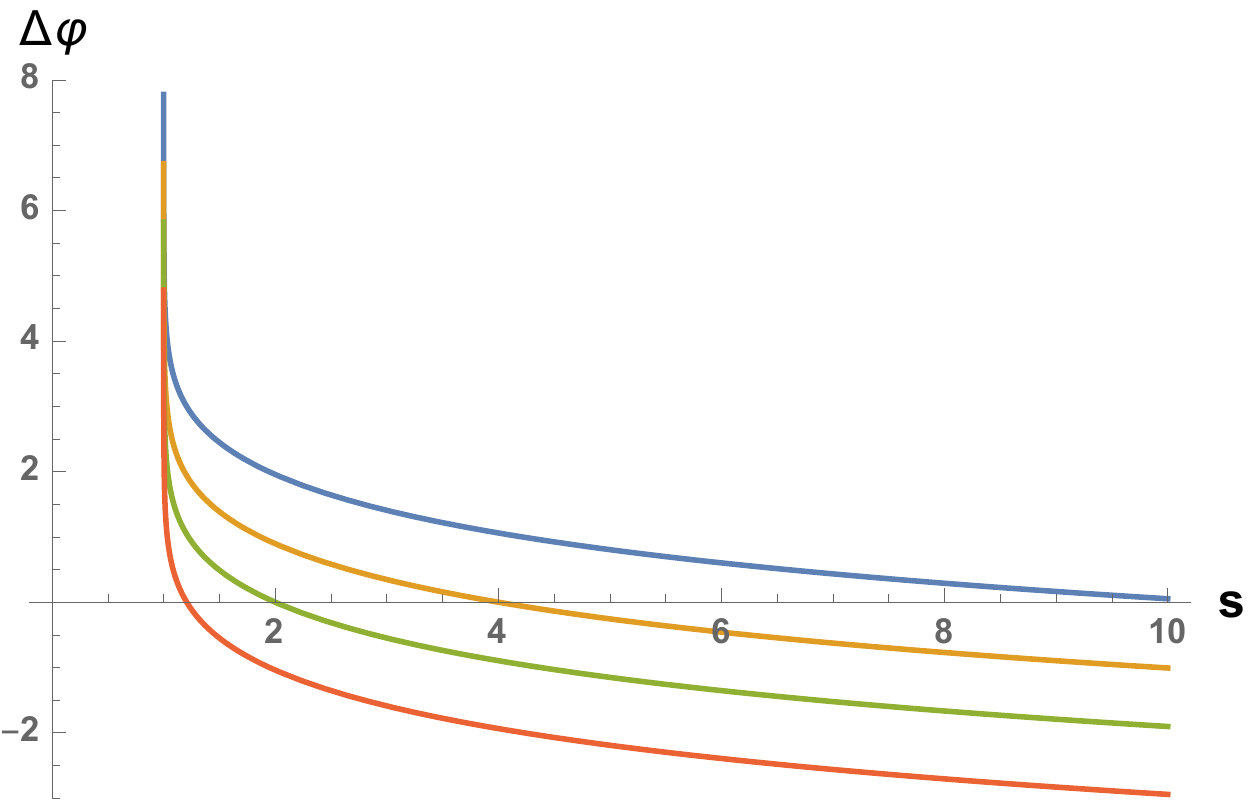}
\includegraphics[width=7.2cm]{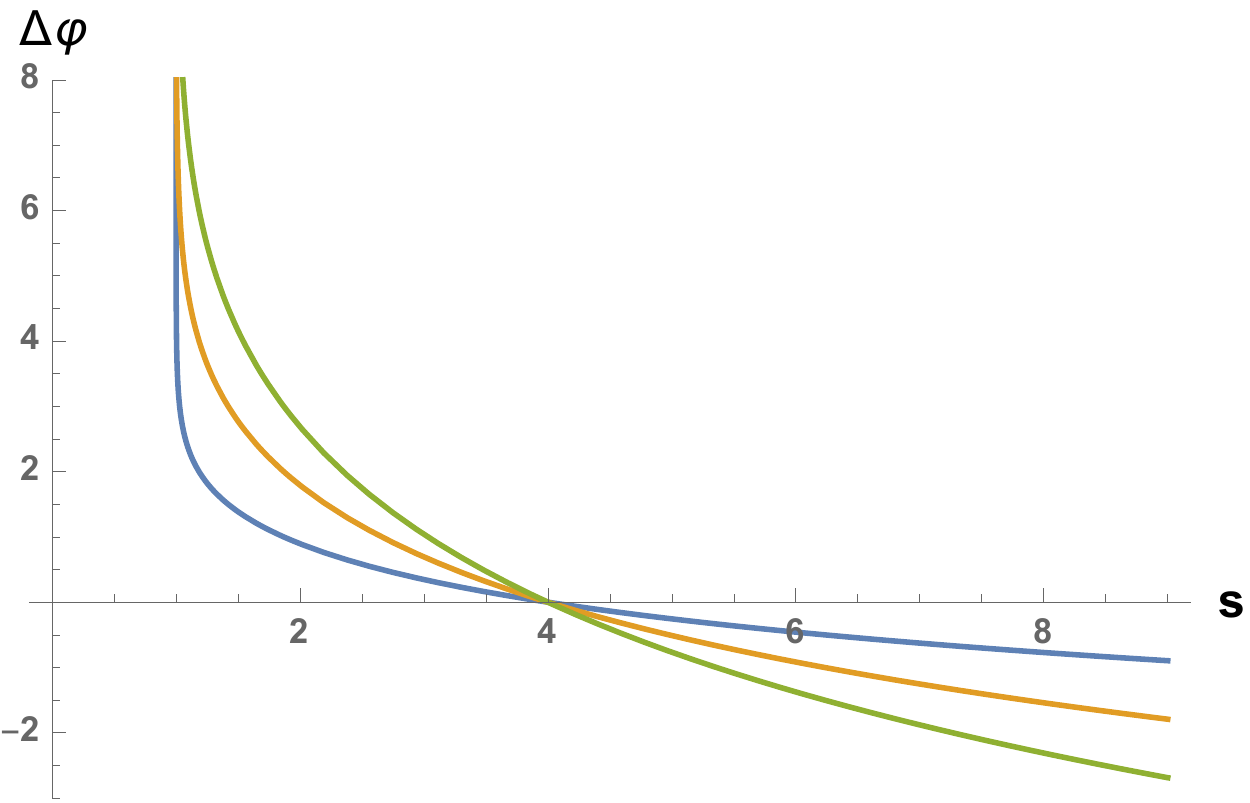}
\caption{Upper bound on the inflatonary field range as a function of the spin of the HS tower. The {\it left panel} shows the behaviour at fixed $\lambda=1$ and for mass ratio $m_0/H=\{0.5, \sqrt{2}, \sqrt{12},10\}$. Higher mass ratio allows for larger field displacement, provided there is cut-off in spin. The {\it right panel} shows the behaviour at fixed mass ratio $m_0/H=\sqrt{12}$ and $\lambda=\{1,1/2,1/3\}$. The maximum cut-off in spin (in this case $s_{\text{max}}<4$) is set just by the mass ratio and it is insensitive to the parameter $\lambda$.}\label{deltaphivss}. 
\end{center}
\vspace{-0.5cm}
\end{figure}

Let us comment on the significance of this result. The bound eq.~\eqref{bound0} strictly depends on  $m_0/H$, that is the ratio between the value of the mass at $\Delta\varphi=0$ and the Hubble scale during inflation. Moreover, each spin $s$ will lead to a different upper bound\footnote{Notice that in this case the upper bound is imposed by a {\it single} HS state, whereas, in the typical picture provided by the SDC, it is the whole infinite tower that effectively imposes the bound through the decay of the quantum gravity cut-off and the consequent breakdown of the EFT.}, which will be more limiting the higher the spin. Whereas a state with spin $s=1$ never forbids any scalar displacement, irrespective of the value of the mass ratio $m_0/H$,  states with higher spins are associated to ever smaller variations in field space (see Fig.~\ref{deltaphivss}). We conclude that the existence of an infinite tower of states of {\it all} spins, for any possible sub- or super-Hubble mass $m_0$, is simply incompatible with inflation (i.e. no scalar field variation is allowed). This situation is shown in Fig.~\ref{deltaphivss} where any line (corresponding to different ratios $m_0/H$) monotonically decreases at large spin.

The only way-out is having a maximum possible value for the spin. In fact, a cut-off in spin $s_{\text{max}}$ would always allow for some scalar field variation provided that
\be
\begin{aligned}
\frac{m_0}{H}>\sqrt{s_{\text{max}}(s_{\text{max}}-1)}\,.
\end{aligned}
\ee
This also implies that, for any cut-off $s_{\text{max}}>1$,  the masses of HS tower should be always super-Hubble with a minimum value of $m_0>\sqrt{2}H$.

Finally, we notice that, while the upper bound on the field range is always very sensitive to the parameter $\lambda$ \cite{Klaewer:2016kiy,Scalisi:2018eaz}, the latter plays no role in determining the cut-off in spin, which can allow for a certain field excursion. The mass ratio is the only relevant parameter which sets the point in spin where $\Delta\varphi=0$ (see right panel of Fig.~\ref{deltaphivss}).

\section{String tower and loophole} \label{Sec:string}

A natural application of the results presented above is provided by the case that the HS tower is identified with the {\it string tower} of states.  The existence of this infinite tower is in fact of fundamental importance for UV completion in string theory. The implications of having string excitations with constant masses in a (quasi-)de Sitter background have been recently discussed in \cite{Noumi:2019ohm,Lust:2019lmq}. However, we know that in certain limits (e.g. at large volume or large dilaton-VEV) the masses of the string states fall exponentially relative to the Planck mass $M_P$.  Interestingly, it has been also noticed that this limit corresponds to an infinite distance point in field space (see e.g. \cite{Lee:2018urn,Font:2019cxq,Lee:2019wij}), thus being another concrete example of the physics predicted by the Swampland Distance Conjecture. This situation offers us a perfect opportunity to apply the above logic and check the physical consequences.

In the case of the string tower, the masses have a relation with the spin given by the Regge trajectory
\be \label{regge}
m^2=s\ M_S^2\,,
\ee
with $M_S$ being the string scale.

The tensionless limit, when the masses decrease exponentially, is therefore fully encoded in the fall-off of the string scale 
\be \label{stringscale}
 M_S= M_S(0) \ e^{-\lambda \Delta\varphi} \qquad \text{as} \qquad\Delta\varphi\rightarrow\infty\,,
\ee
where here large $\Delta\varphi$  can correspond  to limits such as large volume or large dilaton-VEV and $M_S(0)$ is the highest value of the string scale at zero field displacement.

In a (quasi-)de Sitter background, the Higuchi bound will provide a lower bound on the masses, as explained above. For the string tower, the bound reads
\be\label{bound}
\Delta\varphi< \frac{1}{\lambda}\log\left[ \frac{M_{S}(0)}{H}\frac{1}{\sqrt{s-1}}\right]\,.
\ee
Also here, the mere fact that the string tower contains infinite excitations of all spin forbids any scalar field variation, for any ratio $M_{S}(0)/H$. Just a cut-off in spin can allow for some field displacement, analogously to what we discussed in the previous section. In this case, the higher the cut-off in spin, the higher the string scale $M_S(0)$ should be with respect to $H$. For example, a cut-off at $s=5$ implies $M_S(0)>2H$. Since, in principle, one would hope to have a cut-off as high as possible in order to reach an ultraviolet scale, this seems to correspond to impose a huge hierarchy between $M_S(0)$ and $H$. In the following, we show how all these issues can be overcome.

\subsubsection*{Natural cut-off and loophole}

In the case of the string oscillator states, in a (quasi-)de Sitter background, there is in fact a \textit{natural cut-off} beyond which one cannot trust the application of the Higuchi bound (this was pointed out already by \cite{Noumi:2019ohm,Lust:2019lmq} in the case of constant masses). This is obtained by demanding that the length of the string $L\sim \sqrt{s}/M_S$ be smaller than the Hubble radius $R=1/H$. One therefore finds a maximum value for the spin beyond which we cannot trust the Higuchi bound in dS space
\be\label{spin-cut-off}
s_{\text{max}}=\left(\frac{M_S}{H}\right)^2\,.
\ee
This relation implies that the cut-off in spin decreases exponentially while traversing field distances  (and therefore while the string scale drops following eq.~\eqref{stringscale}). The length of a string increases exponentially fast, thus reaching the Hubble radius very soon. If we plug this cut-off into the bound eq.~\eqref{bound}, we notice that we obtain an identity\footnote{We thank Arthur Hebecker for correspondence on this point.} for large values of $s$. For small spin values, the inverse function of eq.~\eqref{spin-cut-off} stays always below the upper bound eq.~\eqref{bound}. We conclude that an exponential decay of the masses of the string tower does not lead to the dire consequences for inflation presented in Sec.~\ref{Sec:spin}, as we have the cut-off in spin which decreases fast enough in field space.

\subsubsection*{UV completion and field range bound}

In order to preserve the conventional argument of UV completion of gravity, we may allow strings with length smaller than the Hubble radius still to have energies reaching at least the Planck scale. The works  \cite{Noumi:2019ohm,Lust:2019lmq} have shown that this provides a bound such as
 \be
M_S>\sqrt{H\ M_P}\,,
\ee
which poses a maximum value on the scale of inflation (the consequences of this bound on brane inflation have been discussed by \cite{Conlon:2019uuy}). Notice that the latter equation corresponds to imposing that the inflationary energy density be below the string scale \cite{Noumi:2019ohm}, a condition that one would plausibly like to assume from the start.

If the string scale $M_S$ drops exponentially as given by eq.~\eqref{stringscale}, we notice that this translates into an upper bound on the field excursion such as
\be\label{boundH}
\Delta\varphi< \frac{1}{\lambda}\log\frac{M_{S}(0)}{\sqrt{H\ M_P}}\,.
\ee
This reduces to the bound of \cite{Scalisi:2018eaz} when $M_{S}(0)=M_P$. For any other initial value of the string scale below the Planck scale, eq.~\eqref{boundH} provides a stronger bound.  Allowing the Hubble parameter $H$ for some field-dependence would certainly relax the bound.

\section{Conclusions}\label{Sec:disc}
In this letter, we have discussed the consequences for inflation of an exponential decay in field space of the masses of an infinite tower of HS states. We have thus shown that preserving unitarity of the theory  \cite{Higuchi:1986py} translates into an upper bound on the inflaton excursion. This allows for some $\Delta\varphi>0$ just if we have a maximum allowed spin in the tower and the masses are super-Hubble. Having {\it all} spins in the tower becomes incompatible with inflation whatsoever.  

This result suggests that if the SDC is a property of quantum gravity, then it should apply just to particles up to spin 1 (in a quasi-de Sitter background), unless they have a cut-off in spins\footnote{Another possibility could be that the SDC is a statement valid in flat space and it requires modifications for backgrounds with curvature.}. The string tower is in fact an example of this situation in string theory. Their cut-off in spin decreases together with the string scale, thus allowing to circumvent entirely the dire implications discussed in Sec.~\ref{Sec:spin}. This argument applies  of course also to other HS towers in string theory that have a similar relation between the spin and the mass  (e.g. wrapped branes). Any mass-independent cut-off would instead lead to a finite field range bound but we do not know of any example in perturbative string theory.  This situation seems to provide some partial endorsement for the recent conjecture of \cite{Lee:2019wij}, which states that an infinite distance limit corresponds either to a decompactification limit (emergence of KK tower) or to some tensionless, weakly-coupled string theory (emergence of tensionless strings). Finally, we have also shown that, whereas exponentially decreasing masses of the string states do not lead to the dire consequences discussed in Sec.~\ref{Sec:spin}, allowing string oscillators in (quasi)-dS to have masses at least up to the Planck scale implies the field range bound eq.~\eqref{boundH}. This is usually stronger than the bound provided by \cite{Scalisi:2018eaz}.

A number of points might deserve further study. We notice that, in the bosonic sector, the spin 2 is the first spin providing a non-trivial bound. It would be interesting to understand the relation to the `spin-2 conjecture' proposed in \cite{Klaewer:2018yxi}. Moreover, one may want to study the deformation of the Regge trajectory for exponentially decaying masses, in the presence of curved backgrounds, along the line of  \cite{Noumi:2019ohm}. We leave these points for future investigations. 

\section*{Acknowledgments}
We would like to thank Arthur Hebecker for enlightening discussions and precious comments. We acknowledge also helpful comments from Matteo Fasiello, Edward Mazenc, Miguel Montero, Hirosi Ooguri, Matthew Reece, Anika Scalisi, Timo Weigand and Alexander Westphal. MS is supported by the Research Foundation Flanders (FWO) and the European Union's Horizon 2020 research and innovation programme under the Marie Sk{\l}odowska-Curie grant agreement No. 665501.

\vspace{1cm}
\bibliographystyle{utphys}
\bibliography{refSDCH}

\providecommand{\href}[2]{#2}\begingroup\raggedright\begin{thebibliography}{10}

\bibitem{Vafa:2005ui}
C.~Vafa, ``{The String landscape and the swampland}'',
\href{http://arxiv.org/abs/hep-th/0509212}{{\tt arXiv:hep-th/0509212
  [hep-th]}}.

\bibitem{Ooguri:2006in}
H.~Ooguri and C.~Vafa, ``{On the Geometry of the String Landscape and the
  Swampland}'', \href{http://dx.doi.org/10.1016/j.nuclphysb.2006.10.033}{{\em
  Nucl. Phys.} {\bf B766} (2007)  21--33},
\href{http://arxiv.org/abs/hep-th/0605264}{{\tt arXiv:hep-th/0605264
  [hep-th]}}.

\bibitem{Brennan:2017rbf}
T.~D. Brennan, F.~Carta, and C.~Vafa, ``{The String Landscape, the Swampland,
  and the Missing Corner}'', \href{http://dx.doi.org/10.22323/1.305.0015}{{\em
  PoS} {\bf TASI2017} (2017)  015},
\href{http://arxiv.org/abs/1711.00864}{{\tt arXiv:1711.00864 [hep-th]}}.

\bibitem{Palti:2019pca}
E.~Palti, ``{The Swampland: Introduction and Review}'',
  \href{http://dx.doi.org/10.1002/prop.201900037}{{\em Fortsch. Phys.} {\bf 67}
  (2019) no.~6, 1900037},
\href{http://arxiv.org/abs/1903.06239}{{\tt arXiv:1903.06239 [hep-th]}}.

\bibitem{Reece19}
M.~Reece,
  ``\href{https://sis-pc15.ulb.ac.be/event/2/contributions/20/attachments/23/61/1_m_reece.pdf}{The
  Swampland Program}''. Review talk at Strings 2019.

\bibitem{Lust:2019zwm}
D.~Lüst, E.~Palti, and C.~Vafa, ``{AdS and the Swampland}'',
  \href{http://dx.doi.org/10.1016/j.physletb.2019.134867}{{\em Phys. Lett. B}
  {\bf 797} (2019)  134867}, \href{http://arxiv.org/abs/1906.05225}{{\tt
  arXiv:1906.05225 [hep-th]}}.

\bibitem{Grimm:2018ohb}
T.~W. Grimm, E.~Palti, and I.~Valenzuela, ``{Infinite Distances in Field Space
  and Massless Towers of States}'',
  \href{http://dx.doi.org/10.1007/JHEP08(2018)143}{{\em JHEP} {\bf 08} (2018)
  143},
\href{http://arxiv.org/abs/1802.08264}{{\tt arXiv:1802.08264 [hep-th]}}.

\bibitem{Blumenhagen:2018nts}
R.~Blumenhagen, D.~Kl{\"a}wer, L.~Schlechter, and F.~Wolf, ``{The Refined
  Swampland Distance Conjecture in Calabi-Yau Moduli Spaces}'',
  \href{http://dx.doi.org/10.1007/JHEP06(2018)052}{{\em JHEP} {\bf 06} (2018)
  052},
\href{http://arxiv.org/abs/1803.04989}{{\tt arXiv:1803.04989 [hep-th]}}.

\bibitem{Lee:2018urn}
S.-J. Lee, W.~Lerche, and T.~Weigand, ``{Tensionless Strings and the Weak
  Gravity Conjecture}'',
\href{http://arxiv.org/abs/1808.05958}{{\tt arXiv:1808.05958 [hep-th]}}.

\bibitem{Lee:2018spm}
S.-J. Lee, W.~Lerche, and T.~Weigand, ``{A Stringy Test of the Scalar Weak
  Gravity Conjecture}'',
  \href{http://dx.doi.org/10.1016/j.nuclphysb.2018.11.001}{{\em Nucl. Phys.}
  {\bf B938} (2019)  321--350},
\href{http://arxiv.org/abs/1810.05169}{{\tt arXiv:1810.05169 [hep-th]}}.

\bibitem{Grimm:2018cpv}
T.~W. Grimm, C.~Li, and E.~Palti, ``{Infinite Distance Networks in Field Space
  and Charge Orbits}'',
\href{http://arxiv.org/abs/1811.02571}{{\tt arXiv:1811.02571 [hep-th]}}.

\bibitem{Buratti:2018xjt}
G.~Buratti, J.~Calderon, and A.~M. Uranga, ``{Transplanckian Axion Monodromy
  !?}'',
\href{http://arxiv.org/abs/1812.05016}{{\tt arXiv:1812.05016 [hep-th]}}.

\bibitem{Gonzalo:2018guu}
E.~Gonzalo, L.~E. Ib{\'a}{\~n}ez, and {\'A}.~M. Uranga, ``{Modular Symmetries
  and the Swampland Conjectures}'',
\href{http://arxiv.org/abs/1812.06520}{{\tt arXiv:1812.06520 [hep-th]}}.

\bibitem{Corvilain:2018lgw}
P.~Corvilain, T.~W. Grimm, and I.~Valenzuela, ``{The Swampland Distance
  Conjecture for Kähler moduli}'',
  \href{http://dx.doi.org/10.1007/JHEP08(2019)075}{{\em JHEP} {\bf 08} (2019)
  075},
\href{http://arxiv.org/abs/1812.07548}{{\tt arXiv:1812.07548 [hep-th]}}.

\bibitem{Font:2019cxq}
A.~Font, A.~Herráez, and L.~E. Ibáñez, ``{The Swampland Distance Conjecture
  and Towers of Tensionless Branes}'',
  \href{http://dx.doi.org/10.1007/JHEP08(2019)044}{{\em JHEP} {\bf 08} (2019)
  044},
\href{http://arxiv.org/abs/1904.05379}{{\tt arXiv:1904.05379 [hep-th]}}.

\bibitem{Lee:2019wij}
S.-J. Lee, W.~Lerche, and T.~Weigand, ``{Emergent Strings from Infinite
  Distance Limits}'',
\href{http://arxiv.org/abs/1910.01135}{{\tt arXiv:1910.01135 [hep-th]}}.

\bibitem{Palti:2015xra}
E.~Palti, ``{On Natural Inflation and Moduli Stabilisation in String Theory}'',
  \href{http://dx.doi.org/10.1007/JHEP10(2015)188}{{\em JHEP} {\bf 10} (2015)
  188},
\href{http://arxiv.org/abs/1508.00009}{{\tt arXiv:1508.00009 [hep-th]}}.

\bibitem{Baume:2016psm}
F.~Baume and E.~Palti, ``{Backreacted Axion Field Ranges in String Theory}'',
  \href{http://dx.doi.org/10.1007/JHEP08(2016)043}{{\em JHEP} {\bf 08} (2016)
  043},
\href{http://arxiv.org/abs/1602.06517}{{\tt arXiv:1602.06517 [hep-th]}}.

\bibitem{Valenzuela:2016yny}
I.~Valenzuela, ``{Backreaction Issues in Axion Monodromy and Minkowski
  4-forms}'', \href{http://dx.doi.org/10.1007/JHEP06(2017)098}{{\em JHEP} {\bf
  06} (2017)  098},
\href{http://arxiv.org/abs/1611.00394}{{\tt arXiv:1611.00394 [hep-th]}}.

\bibitem{Bielleman:2016olv}
S.~Bielleman, L.~E. Ibanez, F.~G. Pedro, I.~Valenzuela, and C.~Wieck,
  ``{Higgs-otic Inflation and Moduli Stabilization}'',
  \href{http://dx.doi.org/10.1007/JHEP02(2017)073}{{\em JHEP} {\bf 02} (2017)
  073},
\href{http://arxiv.org/abs/1611.07084}{{\tt arXiv:1611.07084 [hep-th]}}.

\bibitem{Blumenhagen:2017cxt}
R.~Blumenhagen, I.~Valenzuela, and F.~Wolf, ``{The Swampland Conjecture and
  F-term Axion Monodromy Inflation}'',
  \href{http://dx.doi.org/10.1007/JHEP07(2017)145}{{\em JHEP} {\bf 07} (2017)
  145},
\href{http://arxiv.org/abs/1703.05776}{{\tt arXiv:1703.05776 [hep-th]}}.

\bibitem{Palti:2017elp}
E.~Palti, ``{The Weak Gravity Conjecture and Scalar Fields}'',
  \href{http://dx.doi.org/10.1007/JHEP08(2017)034}{{\em JHEP} {\bf 08} (2017)
  034},
\href{http://arxiv.org/abs/1705.04328}{{\tt arXiv:1705.04328 [hep-th]}}.

\bibitem{Hebecker:2017lxm}
A.~Hebecker, P.~Henkenjohann, and L.~T. Witkowski, ``{Flat Monodromies and a
  Moduli Space Size Conjecture}'',
  \href{http://dx.doi.org/10.1007/JHEP12(2017)033}{{\em JHEP} {\bf 12} (2017)
  033},
\href{http://arxiv.org/abs/1708.06761}{{\tt arXiv:1708.06761 [hep-th]}}.

\bibitem{Cicoli:2018tcq}
M.~Cicoli, D.~Ciupke, C.~Mayrhofer, and P.~Shukla, ``{A Geometrical Upper Bound
  on the Inflaton Range}'',
\href{http://arxiv.org/abs/1801.05434}{{\tt arXiv:1801.05434 [hep-th]}}.

\bibitem{Scalisi:2018eaz}
M.~Scalisi and I.~Valenzuela, ``{Swampland distance conjecture, inflation and
  $\alpha$-attractors}'', \href{http://dx.doi.org/10.1007/JHEP08(2019)160}{{\em
  JHEP} {\bf 08} (2019)  160},
\href{http://arxiv.org/abs/1812.07558}{{\tt arXiv:1812.07558 [hep-th]}}.

\bibitem{Dias:2018ngv}
M.~Dias, J.~Frazer, A.~Retolaza, and A.~Westphal, ``{Primordial Gravitational
  Waves and the Swampland}'',
  \href{http://dx.doi.org/10.1002/prop.201800063}{{\em Fortsch. Phys.} {\bf 67}
  (2019) no.~1-2, 2},
\href{http://arxiv.org/abs/1807.06579}{{\tt arXiv:1807.06579 [hep-th]}}.

\bibitem{Bravo:2019xdo}
R.~Bravo, G.~A. Palma, and S.~Riquelme, ``{A Tip for Landscape Riders:
  Multi-Field Inflation Can Fulfill the Swampland Distance Conjecture}'',
\href{http://arxiv.org/abs/1906.05772}{{\tt arXiv:1906.05772 [hep-th]}}.

\bibitem{Weinberg:1964ew}
S.~Weinberg, ``{Photons and Gravitons in s Matrix Theory: Derivation of Charge
  Conservation and Equality of Gravitational and Inertial Mass}'',
\href{http://dx.doi.org/10.1103/PhysRev.135.B1049}{{\em Phys. Rev.} {\bf 135}
  (1964)  B1049--B1056}.

\bibitem{Coleman:1967ad}
S.~R. Coleman and J.~Mandula, ``{All Possible Symmetries of the S Matrix}'',
\href{http://dx.doi.org/10.1103/PhysRev.159.1251}{{\em Phys. Rev.} {\bf 159}
  (1967)  1251--1256}.

\bibitem{Bekaert:2010hw}
X.~Bekaert, N.~Boulanger, and P.~Sundell, ``{How higher-spin gravity surpasses
  the spin two barrier: no-go theorems versus yes-go examples}'',
  \href{http://dx.doi.org/10.1103/RevModPhys.84.987}{{\em Rev. Mod. Phys.} {\bf
  84} (2012)  987--1009},
\href{http://arxiv.org/abs/1007.0435}{{\tt arXiv:1007.0435 [hep-th]}}.

\bibitem{Vasiliev:1990en}
M.~A. Vasiliev, ``{Consistent equation for interacting gauge fields of all
  spins in (3+1)-dimensions}'',
\href{http://dx.doi.org/10.1016/0370-2693(90)91400-6}{{\em Phys. Lett.} {\bf
  B243} (1990)  378--382}.

\bibitem{Aros:2017ror}
R.~Aros, C.~Iazeolla, J.~Noreña, E.~Sezgin, P.~Sundell, and Y.~Yin, ``{FRW and
  domain walls in higher spin gravity}'',
  \href{http://dx.doi.org/10.1007/JHEP03(2018)153}{{\em JHEP} {\bf 03} (2018)
  153},
\href{http://arxiv.org/abs/1712.02401}{{\tt arXiv:1712.02401 [hep-th]}}.

\bibitem{Singh:1974qz}
L.~P.~S. Singh and C.~R. Hagen, ``{Lagrangian formulation for arbitrary spin.
  1. The boson case}'',
\href{http://dx.doi.org/10.1103/PhysRevD.9.898}{{\em Phys. Rev.} {\bf D9}
  (1974)  898--909}.

\bibitem{Singh:1974rc}
L.~P.~S. Singh and C.~R. Hagen, ``{Lagrangian formulation for arbitrary spin.
  2. The fermion case}'',
\href{http://dx.doi.org/10.1103/PhysRevD.9.910}{{\em Phys. Rev.} {\bf D9}
  (1974)  910--920}.

\bibitem{Zinoviev:2001dt}
{\relax Yu}.~M. Zinoviev, ``{On massive high spin particles in AdS}'',
\href{http://arxiv.org/abs/hep-th/0108192}{{\tt arXiv:hep-th/0108192
  [hep-th]}}.

\bibitem{Arkani-Hamed:2015bza}
N.~Arkani-Hamed and J.~Maldacena, ``{Cosmological Collider Physics}'',
\href{http://arxiv.org/abs/1503.08043}{{\tt arXiv:1503.08043 [hep-th]}}.

\bibitem{Lee:2016vti}
H.~Lee, D.~Baumann, and G.~L. Pimentel, ``{Non-Gaussianity as a Particle
  Detector}'', \href{http://dx.doi.org/10.1007/JHEP12(2016)040}{{\em JHEP} {\bf
  12} (2016)  040},
\href{http://arxiv.org/abs/1607.03735}{{\tt arXiv:1607.03735 [hep-th]}}.

\bibitem{Kehagias:2017cym}
A.~Kehagias and A.~Riotto, ``{On the Inflationary Perturbations of Massive
  Higher-Spin Fields}'',
  \href{http://dx.doi.org/10.1088/1475-7516/2017/07/046}{{\em JCAP} {\bf 1707}
  (2017) no.~07, 046},
\href{http://arxiv.org/abs/1705.05834}{{\tt arXiv:1705.05834 [hep-th]}}.

\bibitem{Alexander:2019vtb}
S.~Alexander, S.~J. Gates, L.~Jenks, K.~Koutrolikos, and E.~McDonough,
  ``{Higher Spin Supersymmetry at the Cosmological Collider: Sculpting SUSY
  Rilles in the CMB}'', \href{http://dx.doi.org/10.1007/JHEP10(2019)156}{{\em
  JHEP} {\bf 10} (2019)  156},
\href{http://arxiv.org/abs/1907.05829}{{\tt arXiv:1907.05829 [hep-th]}}.

\bibitem{Higuchi:1986py}
A.~Higuchi, ``{Forbidden Mass Range for Spin-2 Field Theory in De Sitter
  Space-time}'',
\href{http://dx.doi.org/10.1016/0550-3213(87)90691-2}{{\em Nucl. Phys.} {\bf
  B282} (1987)  397--436}.

\bibitem{Noumi:2019ohm}
T.~Noumi, T.~Takeuchi, and S.~Zhou, ``{String Regge trajectory on de Sitter
  space and implications to inflation}'',
\href{http://arxiv.org/abs/1907.02535}{{\tt arXiv:1907.02535 [hep-th]}}.

\bibitem{Lust:2019lmq}
D.~Lüst and E.~Palti, ``{A Note on String Excitations and the Higuchi
  Bound}'',
\href{http://arxiv.org/abs/1907.04161}{{\tt arXiv:1907.04161 [hep-th]}}.

\bibitem{Klaewer:2016kiy}
D.~Klaewer and E.~Palti, ``{Super-Planckian Spatial Field Variations and
  Quantum Gravity}'',
\href{http://arxiv.org/abs/1610.00010}{{\tt arXiv:1610.00010 [hep-th]}}.

\bibitem{Heidenreich:2018kpg}
B.~Heidenreich, M.~Reece, and T.~Rudelius, ``{Emergence of Weak Coupling at
  Large Distance in Quantum Gravity}'',
  \href{http://dx.doi.org/10.1103/PhysRevLett.121.051601}{{\em Phys. Rev.
  Lett.} {\bf 121} (2018) no.~5, 051601},
\href{http://arxiv.org/abs/1802.08698}{{\tt arXiv:1802.08698 [hep-th]}}.

\bibitem{Hebecker:2018vxz}
A.~Hebecker and T.~Wrase, ``{The Asymptotic dS Swampland Conjecture ‐ a
  Simplified Derivation and a Potential Loophole}'',
  \href{http://dx.doi.org/10.1002/prop.201800097}{{\em Fortsch. Phys.} {\bf 67}
  (2019) no.~1-2, 1800097},
\href{http://arxiv.org/abs/1810.08182}{{\tt arXiv:1810.08182 [hep-th]}}.

\bibitem{Dvali:2007wp}
G.~Dvali and M.~Redi, ``{Black Hole Bound on the Number of Species and Quantum
  Gravity at LHC}'', \href{http://dx.doi.org/10.1103/PhysRevD.77.045027}{{\em
  Phys. Rev.} {\bf D77} (2008)  045027},
\href{http://arxiv.org/abs/0710.4344}{{\tt arXiv:0710.4344 [hep-th]}}.

\bibitem{Dvali:2007hz}
G.~Dvali, ``{Black Holes and Large N Species Solution to the Hierarchy
  Problem}'', \href{http://dx.doi.org/10.1002/prop.201000009}{{\em Fortsch.
  Phys.} {\bf 58} (2010)  528--536},
\href{http://arxiv.org/abs/0706.2050}{{\tt arXiv:0706.2050 [hep-th]}}.

\bibitem{Akrami:2018odb}
{\bf Planck} Collaboration, Y.~Akrami {\em et al.}, ``{Planck 2018 results. X.
  Constraints on inflation}'',
\href{http://arxiv.org/abs/1807.06211}{{\tt arXiv:1807.06211 [astro-ph.CO]}}.

\bibitem{Aragam:2019khr}
V.~Aragam, S.~Paban, and R.~Rosati, ``{Multi-field Inflation in High-Slope
  Potentials}'',
\href{http://arxiv.org/abs/1905.07495}{{\tt arXiv:1905.07495 [hep-th]}}.

\bibitem{Conlon:2019uuy}
J.~P. Conlon, ``{A Note on Brane Inflation}'',
\href{http://arxiv.org/abs/1911.04776}{{\tt arXiv:1911.04776 [hep-th]}}.

\bibitem{Klaewer:2018yxi}
D.~Klaewer, D.~Lüst, and E.~Palti, ``{A Spin‐2 Conjecture on the
  Swampland}'', \href{http://dx.doi.org/10.1002/prop.201800102}{{\em Fortsch.
  Phys.} {\bf 67} (2019) no.~1-2, 1800102},
\href{http://arxiv.org/abs/1811.07908}{{\tt arXiv:1811.07908 [hep-th]}}.

\end{thebibliography}\endgroup

\end{document}